\documentclass[aps,pra,twocolumn,groupedaddress,showpacs,10pt]{revtex4-1}

\usepackage{graphicx}
\usepackage{dcolumn}
\usepackage{bm}
\usepackage{setspace}
\usepackage{color}
\usepackage{overpic}
\usepackage{sidecap}
\usepackage{braket}
\usepackage{amsmath}

\usepackage[english]{babel}
\draft

\begin{document}

\title{Observation of a Spinning Top in a Bose-Einstein Condensate}

\author{R.~N.~Bisset$^1$}
\email[]{rnbisset@gmail.com}
\author{S.~Serafini$^1$}
\author{E.~Iseni$^1$}
\author{M.~Barbiero$^{2,1}$}
\author{T.~Bienaim\'e$^1$}
\author{G.~Lamporesi$^1$}
\author{G.~Ferrari$^1$}
\author{F.~Dalfovo$^1$}

\affiliation{$^1$INO--CNR BEC Center and Dipartimento di Fisica, Universit\`a di Trento, 38123 Povo, Italy}
\affiliation{$^2$Politecnico di Torino, Dipartimento di Elettronica e Telecomunicazioni, Corso duca degli Abruzzi 24, 10129 Torino, Italy}

\begin{abstract}

Boundaries strongly affect the behavior of quantized vortices in Bose-Einstein condensates, 
a phenomenon particularly evident in elongated cigar-shaped traps where vortices tend to orient 
along a short direction to minimize energy. Remarkably, contributions to the angular momentum 
of these vortices are tightly confined to the region surrounding the core, in stark contrast to 
untrapped condensates where all atoms contribute $\hbar$. We develop a theoretical model and 
use this, in combination with numerical simulations, to show that such localized vortices precess 
in an analogous manner to that of a classical spinning top. We experimentally verify this 
spinning-top behavior with our real-time imaging technique that allows for the tracking of position 
and orientation of vortices as they dynamically evolve. Finally, we perform an in-depth numerical 
investigation of our real-time expansion and imaging method, with the aim of guiding future 
experimental implementation, as well as outlining directions for its improvement.

\end{abstract}

\maketitle

\section{Introduction}

Bose-Einstein condensates (BEC) are ideally suited for the study of quantum vortices, owning to their purity and high-degree of tunability \cite{Fetter09}, and this inherent flexibility has inspired experimental and theoretical works in a wide variety of settings.
Vortex lattices provide the fundamental means for bulk superfluid flow in rotating BECs \cite{Madison00,Hodby2001,AboShaeer01,Engels03} while, on the other hand, vortices also lie at the heart of quantum turbulence in nonequilibrium systems \cite{Kozik04,Kozik06,Henn09,Neely2013,Kwon14,Baggaley12}.
Boundaries play a central role, and when a vortex line pierces a condensate's surface it does so at an angle perpendicular to it.
When a vortex is positioned off-center, it tends to orbit around the condensate center at an increasing frequency as it spirals outward due to dissipation \cite{Anderson00,Rokhsar1997,Serafini15}.
Under the influence of a pancake-shaped trapping potential, vortices tend to minimize their energy by aligning along the short direction and, at finite temperature, the vortex-unbinding Berezinskii-Kosterlitz-Thouless phase transition was studied \cite{Hadzibabic2006,Clade2009,Tung2010,Hung2011,Holzmann2008,Bisset2009}.
In three dimensions, in addition to vortex lines \cite{Matthews99,Rosenbusch02}, vortices can fold to create rings \cite{Anderson01,Shomroni09,Ginsberg2005,Becker13,Komineas2007,Kevrekidis2015} and even more exotic structures like 
hopfions \cite{Kartashov2014,Bidasyuk2015,Bisset2015} and Chladni solitons \cite{Mateo14}.
Spiralling undulations of the cores, known as Kelvin waves, are responsible for the so-called Kelvin-wave cascade \cite{Kozik04,Kozik06}.

\begin{figure}[t!]
\includegraphics[width=1\columnwidth]{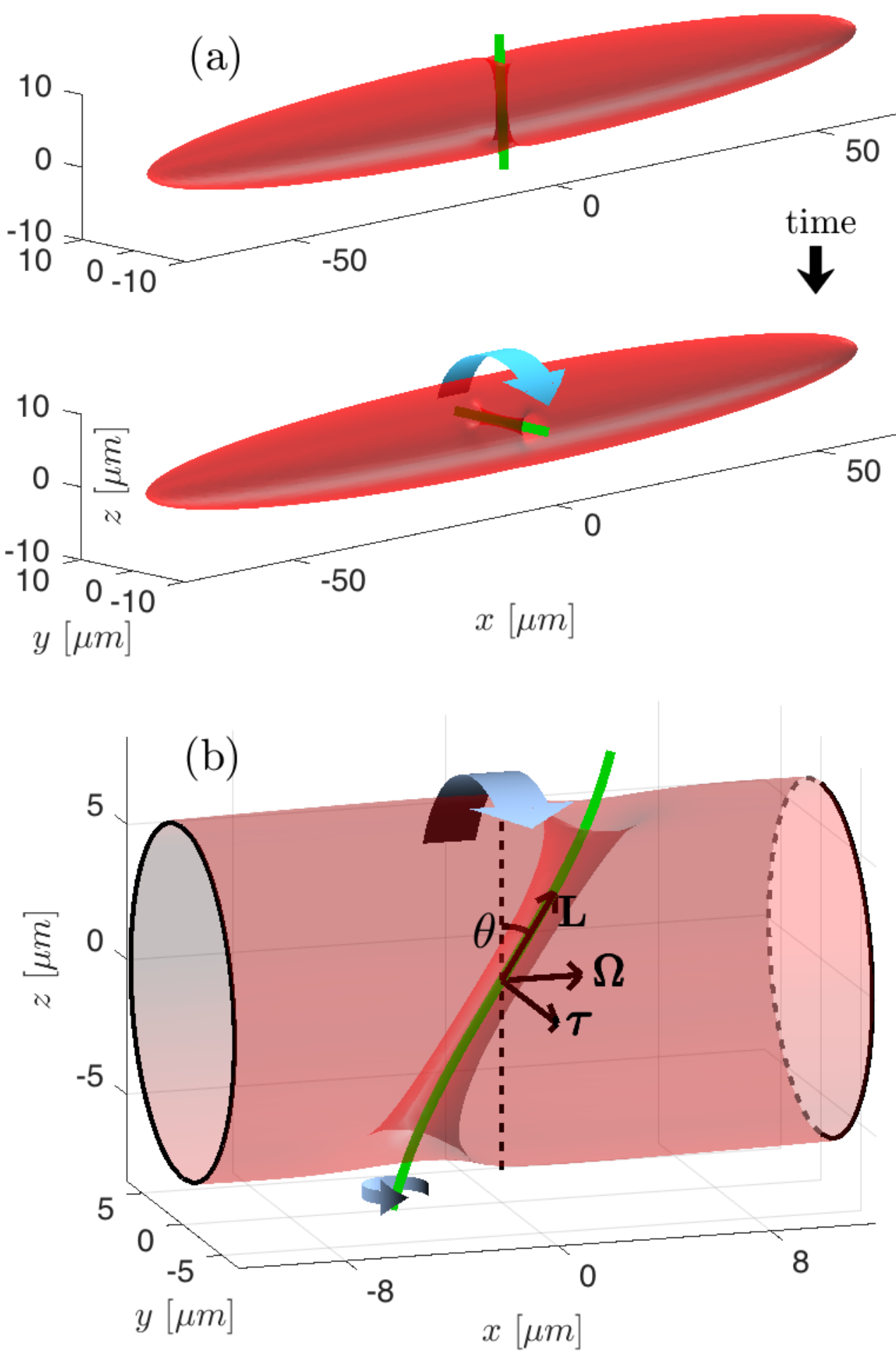} 
\caption{Illustration of solitonic vortex precession. The shaded region represents the isosurface at half the peak density of the condensate, while the green lines indicate the vortex core. (a) An example of precession around the $x$ axis by a quarter of a period, changing from a vertical to a horizontal orientation. On this scale the axial tilt is too small to be visible. (b)
Condensate segment containing a tilted solitonic vortex whose core lies in the $y=0$ plane with a tilt angle $\theta$ into the long ($x$) axis of the trap [the tilt angle is larger here than for the vortex in (a)]. The dashed line is a vertical reference.
The vortex experiences a torque $\boldsymbol\tau$ that acts to reduce the core length, i.e., by attempting to reduce $\theta$. An arrow indicating the direction of $\boldsymbol\tau$, as well as arrows for the directions of the angular momentum of the vortex $\bf L$ and precession $\boldsymbol\Omega$, are shown. The senses of the vortex and the precession are indicated by the bottom and top blue arrows, respectively.
} 
\label{Fig:1}
\end{figure}

In three-dimensional (3D) cigar-shaped traps the most stable defect is the so-called solitonic vortex, a short vortex line that pierces the condensate through its side \cite{Tsatsos2016,Brand02,Komineas03,Brand01,Becker13}. While they are indeed vortices, solitonic vortices possess some solitonic characteristics such as being more localized and, on the coarse-grain scale, they cause a $\pi$ phase jump between each end of the cigar which results in a planar density depletion after expansion \cite{Tylutki15}. Solitonic vortices, which were recently realized in experiments with bosons \cite{Donadello14,Becker13} and fermions \cite{Ku14}, tend to be long-lived and orbit about the condensate center on an elliptical path, along which the core remains surrounded by a roughly constant density. 

We experimentally produce solitonic vortices in cigar-shaped traps that are inherited from the condensate formation process, thanks to the Kibble-Zurek mechanism \cite{LamporesiKZM13,Donadello16} (also see \cite{Kibble76,Zurek85,Weiler08,Freilich10,Corman14,Navon15,Chomaz15}).  A BEC is formed by a cooling quench across the transition temperature, where a symmetry-breaking phase transition occurs. If the quench is fast enough, distant regions of the system do not have sufficient time to communicate and hence they randomly develop order parameters disparate from one other. Defects, such as dark solitons or quantized vortices, then become trapped at the boundaries between such regions. With an appropriate imaging technique we track the axial position and the orientation of the vortex lines which remain in our BECs, as remnants of the Kibble-Zurek mechanism and the subsequent post-quench dynamics. 

Some of these vortices exhibit a peculiar rotation of their core around the long axis of the trap as depicted in Fig.~\ref{Fig:1} (a). In this work we show that such a rotation is caused by a tilt of the vortex line out of the radial plane and towards the symmetry axis, as shown in Fig.~\ref{Fig:1} (b); the tilt implies an increase of the vortex line length, with a consequent energy cost and an induced torque. The torque produces the precession of the vortex around the axial direction, in an analogous manner to a classical spinning top.  The analogy works well because the solitonic vortex is a localized object, in contrast to regular 3D vortices. 

We verify this spinning top behavior by performing numerical simulations, using the Gross-Pitaevskii equation (GPE). We find that a solution of the GPE exists corresponding to a tilted vortex, which is stationary in a reference frame rotating around the long axis of the trap. We then use such a stationary state as an input of real time GP simulations in the nonrotating BEC and we observe that the vortex line keeps rotating at a constant angular velocity. We use the GPE also to simulate the extraction and expansion of atoms as performed in the experiments, in order to reproduce our minimally-destructive imaging scheme that is able to track the orientation and position of the spinning vortices in real time.

The paper is structured as follows: In Sec.~\ref{Sec:SpinTheory} we present our spinning top theory for the solitonic vortex. In Sec.~\ref{Sec:Numerics}, we outline our numerical approach for simulating real-time dynamics of 3D cigar-shaped condensates. We also explain how to obtain solitonic vortex initial states for our precession simulations. 
The focus of Sec.~\ref{Sec:InTrapRes} is the comparison of our spinning top model with our numerical results; in particular, we calculate the precession frequency versus the axial tilt angle $\theta$ (see Fig.~\ref{Fig:1} (b) for the definition of $\theta$).
In Sec.~\ref{Sec:ExtProc} we outline our experimental extraction procedure and provide details for how this is numerically simulated. Section \ref{Sec:ExtRes} presents our experimental observations of solitonic vortex spinning tops, alongside their numerical counterpart for comparison. We also discuss our extraction and imaging scheme, and suggest directions for its improvement.
We conclude with section \ref{Sec:Conc}.

\section{Formalism}

\subsection{Theory of the BEC spinning top}\label{Sec:SpinTheory}


\begin{figure*}[t!]
\includegraphics[width=0.7\textwidth]{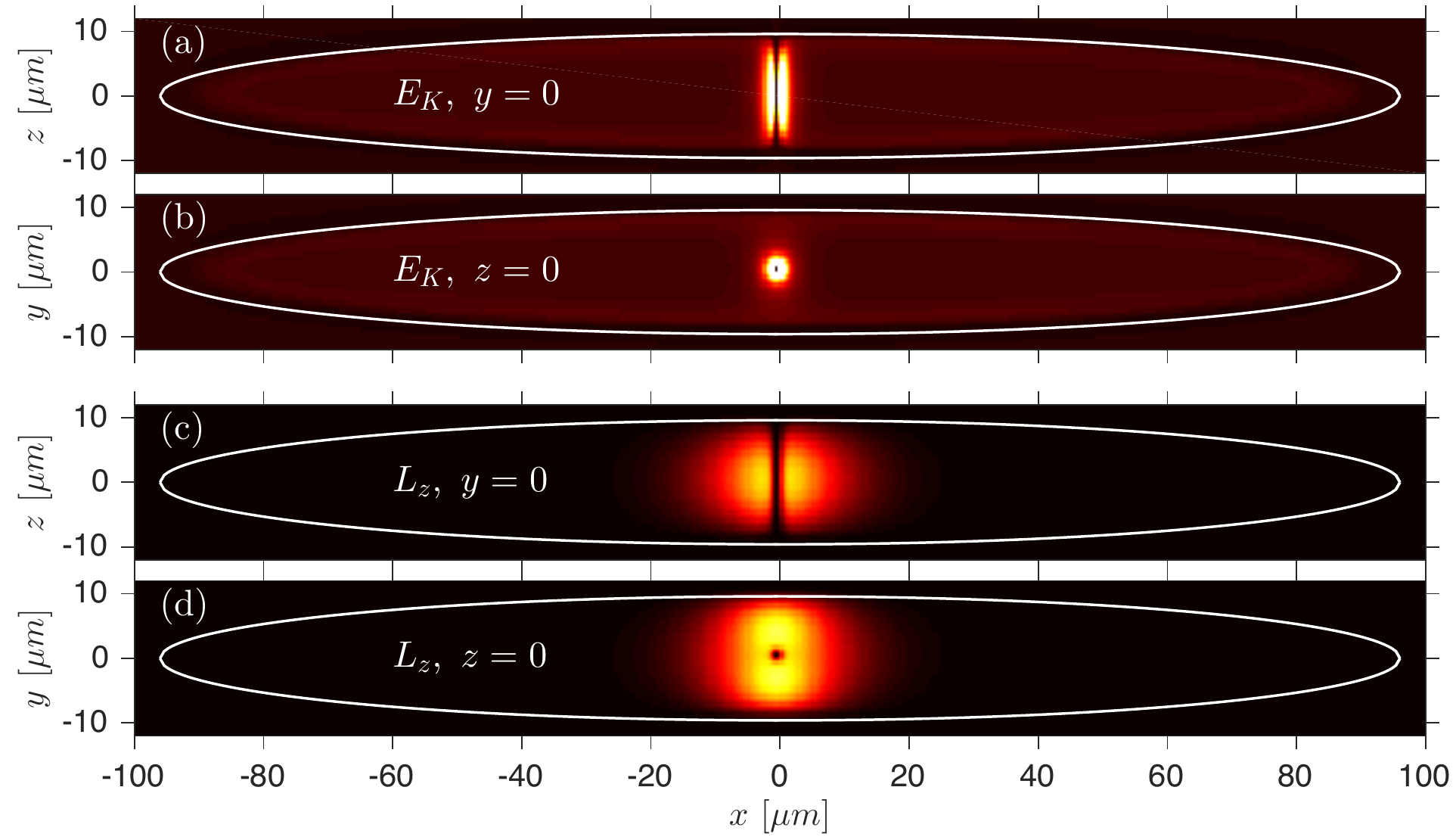}
\caption{Kinetic energy (a)(b) and angular momentum (c)(d) density of a solitonic-vortex stationary state in a nonrotating BEC. The vortex core is aligned along the $z$ axis, and we consider both parallel and perpendicular slices.
Due to symmetry, only the $z$-component gives a non-zero contribution to the total angular momentum, and here we plot its real part, i.e.~$L_z(\mathbf{x}) = \Re\{\psi^*(\mathbf{x}) \hat{L}_z \psi(\mathbf{x}) \}$ where $\hat{L}_z=\hbar/i(x\partial_y - y\partial_x)$. The white lines denote where the Thomas-Fermi density is expected to vanish. This data is produced by a GPE simulation of $8\times 10^5$ $^{23}$Na atoms with $\omega_\perp/\omega_z=10$ and $\omega_\perp/2\pi=92$ Hz, which gives $\mu/\hbar\omega_\perp=9.72$.
}
\label{Fig:LnE}
\end{figure*}

In contrast to a vortex in an untrapped system the solitonic vortex is a highly localized object.
This local character is what allows for a close analogy with the classical spinning top, and serves as the basis for our analytic approach.

For an isolated straight vortex line in an untrapped condensate, each atom contributes $\hbar$ to the angular momentum regardless of the distance from the core.
Consequently, the angular momentum per unit length rapidly diverges with increasing system size. Realistic trapped systems, however, offer qualitative differences: close boundaries in anisotropic traps act to restrict the superfluid flow, which limits the fraction of atoms that contribute to the angular momentum to be only those in the vicinity of the core.
A solitonic vortex in a cigar-shaped condensate is an excellent example, and Figs.~\ref{Fig:LnE} (c) and (d) show that the dominant contributions to the angular momentum are tightly localized about the core.
Furthermore, as can be seen in Figs.~\ref{Fig:LnE} (a) and (b), the kinetic energy density is also localized about the core. 

One way for a vortex to decrease its energy is for its core length to shorten.
It follows, then, that a solitonic vortex that is tilted by an angle $\theta$ into the long axis of the trap, as shown in Fig.~\ref{Fig:1} (b), experiences a restoring torque that acts to reduce $\theta$.
The torque $\boldsymbol\tau$ modifies the angular momentum $\mathbf{L}$ according to
\begin{equation}
\boldsymbol\tau = \frac{d\mathbf{L}}{dt} = \boldsymbol\Omega \times \mathbf{L} \, ,
\end{equation}
causing a precession $\boldsymbol\Omega$ about the $x$ axis, as illustrated in Fig.~\ref{Fig:1}.
Note that we assume that the angular momentum associated with the precession is much smaller than the angular momentum of the vortex itself.
The angular precession frequency, in radians per second, is then given by
\begin{equation}
\Omega \equiv |\boldsymbol{\Omega}| = \frac{\tau}{L\cos\theta} \label{Eq:PrecessF} \, ,
\end{equation}
where $\tau = |\boldsymbol\tau|$ and $L=|\bf{L}|$.

The torque $\tau$ can be calculated as follows.
The energy of an untilted solitonic vortex in a highly-elongated condensate ($\omega_\perp \gg \omega_x$, for radial and axial harmonic trapping frequencies, respectively) has been calculated \cite{Ku14,Mateo2015}, to logarithmic accuracy, to be
\begin{equation}
E^0 =  \frac{4}{3} \frac{\pi n_0\hbar^2 R_\perp}{m} \ln{\left(\frac{R_\perp}{\xi}\right)} \, ,
\label{EQ:SVE}
\end{equation}
with $\xi=\hbar/\sqrt{2m\mu}$ being the healing length, $R_\perp = \sqrt{2\mu/m\omega_\perp^2}$ is the Thomas-Fermi radius in the tight-confinement direction, $n_0$ is the peak density and $m$ is the mass.
Tilting the solitonic vortex so that the core develops a nonzero axial ($x$) component, as shown in Fig.~\ref{Fig:1} (b), increases its energy. To lowest order in $\theta$ this is described by
$E = E^0/\cos\theta$,
which produces a torque of strength
\begin{equation}
\tau = \frac{dE}{d\theta} = \frac{4}{3} A_E \frac{\pi n_0\hbar^2 R_\perp}{m} \ln{\left(\frac{R_\perp}{\xi}\right)} \frac{\sin\theta}{\cos^2\theta} \,. \label{Eq:tau}
\end{equation}
The constant $A_E$, to be determined numerically by solving the GPE, is a correction factor to Eq.~(\ref{EQ:SVE}) and is expected to be approximately unity.
Its purpose is, both, a way to quantify the accuracy of Eq.~(\ref{EQ:SVE}) for realistic chemical potentials and 3D trapping, and to improve the value of $\tau$ for the prediction of the precession frequency given by Eq.~(\ref{Eq:PrecessF}).
For the regime considered in our simulations (i.e., $\mu/\hbar\omega_\perp=9.72$ and $\omega_\perp/\omega_x=10$) we find $A_E = 0.944$, which is indeed close to unity.\\

As shown in Fig.~\ref{Fig:LnE} (c) and (d), the atoms contributing to the angular momentum are predominantly confined to the vicinity of the core, within a radius $R_\perp$. If each of these atoms contributes approximately $\hbar$ then this gives a total angular momentum,
\begin{equation}
L = A_L \pi  R_\perp^3 n_0 \hbar \, ,
\label{Eq:L}
\end{equation}
where $A_L$ is some constant, of order unity.
We also calculate this numerically and find that for the regime of our simulations $A_L = 0.995$.

Finally, by substituting Eqs.~(\ref{Eq:tau}) and (\ref{Eq:L}) into Eq.~(\ref{Eq:PrecessF}), we obtain a prediction for the precession frequency,
\begin{align}
\frac{\Omega_{\rm A}}{\omega_\perp} &= \frac{A_E}{A_L} \left[ \frac{2\ln(2\tilde\mu)}{3\tilde\mu} \right] \frac{\sin\theta}{\cos^3\theta} \\
&\approx \frac{A_E}{A_L} \left[ \frac{2\ln(2\tilde\mu)}{3\tilde\mu} \right] \theta  \,, \label{Eq:PrecessAna}
\end{align}
which is a function of the tilt angle $\theta$
and the dimensionless chemical potential $\tilde\mu = \mu/\hbar\omega_\perp$.

\subsection{Numerics}\label{Sec:Numerics}

Our analytic predictions are supported by full numerical simulations of the time-dependent GPE \cite{Pitaevskii16},
\begin{equation}
i\hbar\frac{\partial\psi(\mathbf{x})}{\partial t} = \left[-\frac{\hbar^2}{2m}\nabla^2 + V(\mathbf{x}) + g|\psi(\mathbf{x})|^2\right]\psi(\mathbf{x}) , \label{Eq:GPE}
\end{equation}
where interactions are characterized by $g=4\pi\hbar^2a_s/m$, with $a_s$ being the $s$-wave scattering length.
We consider a 3D harmonic trapping potential,
\begin{equation}
V(\mathbf{x})=\frac{1}{2}m\omega_x^2x^2 + \frac{1}{2}m\omega_y^2y^2 + \frac{1}{2}m\omega_z^2z^2 ,
\end{equation}
that is cylindrically symmetric and elongated along the $x$ direction, i.e.~$\omega_x \ll \omega_{y,z} =\omega_\perp$.

Initial states, for subsequent precession dynamics, are created by making use of the rotating-trap GPE \cite{Castin1999,Aftalion01,Pitaevskii16},
\begin{equation}
\mu\psi(\mathbf{x}) = \left[-\frac{\hbar^2}{2m}\nabla^2 + V(\mathbf{x}) + g|\psi(\mathbf{x})|^2 - \Omega_{\mathrm{tr}} \hat L_x\right]\psi(\mathbf{x}) , \label{Eq:GPErot}
\end{equation}
where $\Omega_{\mathrm{tr}}$ is the trap-rotation frequency and $\hat{L}_x = \hbar/i(y\partial_z-z\partial_y)$, so that the axis of rotation is coincident with the long ($x$) axis.
The procedure begins by imprinting an untilted solitonic vortex onto the ground state of the GPE [Eq.~(\ref{Eq:GPE})].
A tilted solitonic-vortex stationary state is then obtained by evolving this state according to Eq.~(\ref{Eq:GPErot}) with imaginary time evolution; the adjustment of $\Omega_{\mathrm{tr}}$ acts as a control knob for the tilt angle $\theta$.

For the purpose of investigating the in-trap dynamics of precessing solitonic vortices we consider $N = 8\times 10^5$ $^{23}$Na atoms in a cigar-shaped trap, having $\omega_\perp = 10\omega_x$. The scattering length is 54.54(20)\,$a_0$, for Bohr radius $a_0$ \cite{Knoop2011}, and the radial trapping frequency is given by  $\omega_\perp/2\pi=92$\,Hz which then corresponds to $\mu/\hbar\omega_\perp=9.72$, a system well within the Thomas-Fermi regime.
Time propagation is performed with a 4th order Runge-Kutta integration method and a time step size of 1.7\,$\mu$s.
The 3D numerical grid for the simulation of in-trap dynamics has size $\{L_x,L_y,L_z\}=\{229,34.9,34.9\}\mu$m, and there are $\{N_x,N_y,N_z\}=\{600,60,60\}$ points in the respective directions.
The grid has linear spacing and we employ fast Fourier transforms to evaluate the kinetic energy terms at each time step.

\section{Results}
\subsection{In-Trap Behavior: Numerics and Analytics}\label{Sec:InTrapRes}

\begin{figure}
\includegraphics[width=1\columnwidth]{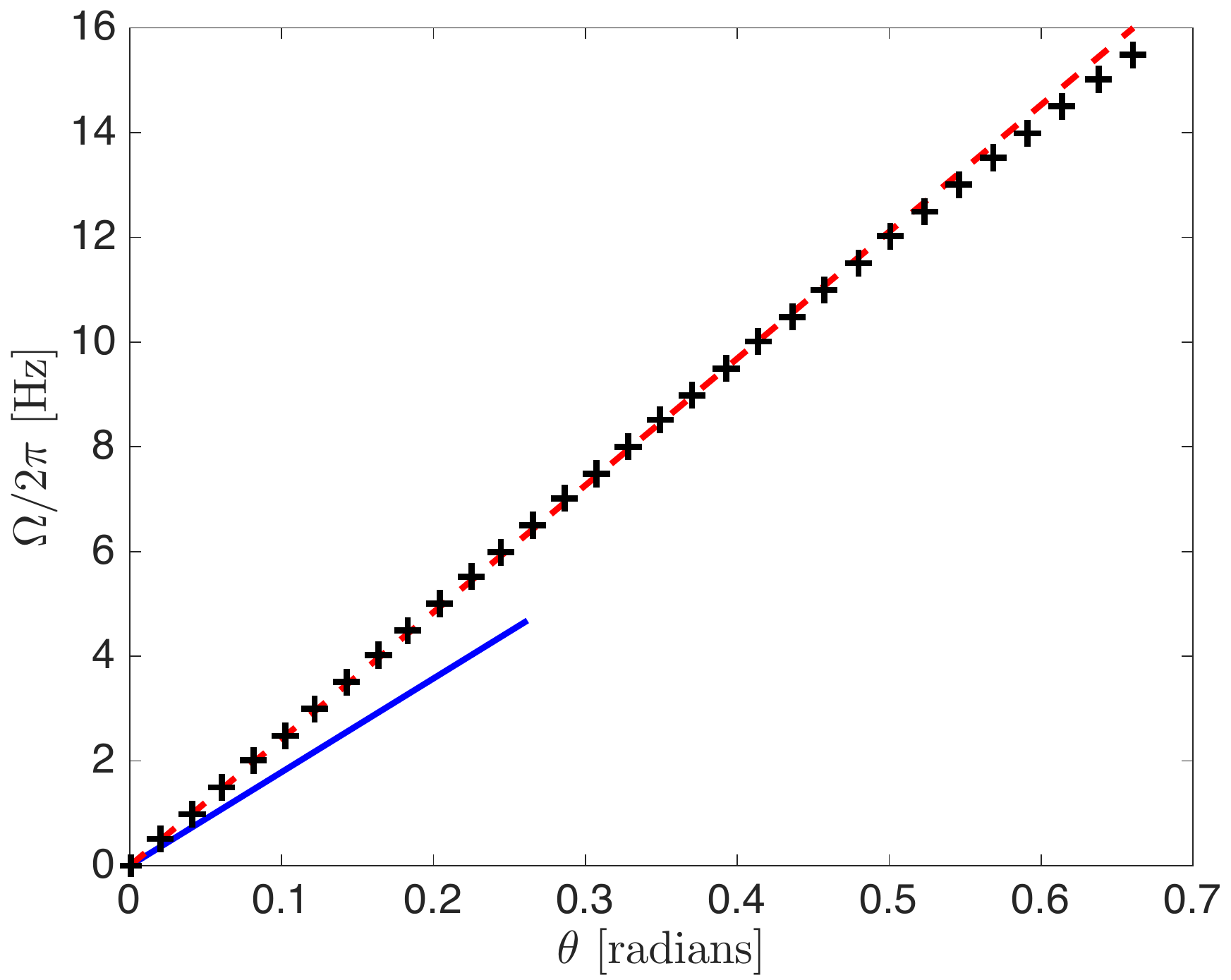}
\caption{Precession frequency versus tilt angle for a solitonic vortex.
Full numerical results (plus signs) and the analytic prediction $\Omega_{\rm A}$ (solid line) provided by Eq.~(\ref{Eq:PrecessAna}) are displayed. The red dashed line is a straight line to guide the eye, having a slope that matches the first two numerical data points. Note that the chemical potential used in Eq.~(\ref{Eq:PrecessAna}) was chosen to match that of the numerical simulations. Incidentally, the difference in chemical potential due to the presence or absence of a vortex has no discernible effect on the results presented here. For Eq.~(\ref{Eq:PrecessAna}) we use the numerically  determined adjustment factors $A_E=0.944$ and $A_L=0.995$ obtained from the GPE (see main text). Parameters are the same as in Fig.~\ref{Fig:LnE}.
} 
\label{Fig:Precess}
\end{figure}

Recall from Fig.~\ref{Fig:1} (b) that a solitonic vortex that is tilted into the long ($x$) axis of the trap will feel a torque and precess about this axis without changing its shape;
in this part we numerically investigate how the precession frequency $\Omega$ depends on the tilt angle $\theta$.
In Sec.~\ref{Sec:Numerics} we outlined how to construct tilted solitonic vortex states by solving the rotating-trap GPE [Eq.~(\ref{Eq:GPErot})] and using its stationary states.
It is a relatively straightforward extension to turn off the trap rotation and then to evolve these initial states in real time, i.e.~by solving Eq.~(\ref{Eq:GPE}). In fact, the resulting real-time precession frequencies $\Omega$ were compared with the corresponding values of $\Omega_{\rm tr}$, used for initial state preparation, as a means to check the convergence.
The numerical data are presented in Fig.~\ref{Fig:Precess} as plus symbols. As expected from our analytic model [Eq.~(\ref{Eq:PrecessAna})], for small tilt angles the precession frequency exhibits a linear dependence although, remarkably, this nearly-linear relationship extends up to around $\theta$ = 0.6 radians.
The agreement with the analytic prediction given by Eq.~(\ref{Eq:PrecessAna}) (solid line) is also quantitatively reasonable, with the analytic prediction being around 27\% smaller.
It should be noted that this discrepancy is not entirely surprising given that the system is not a rigid body and, as the vortex tilts, the superfluid flow has to contend with the anisotropic boundary.\\

\begin{figure}
\includegraphics[width=1\columnwidth]{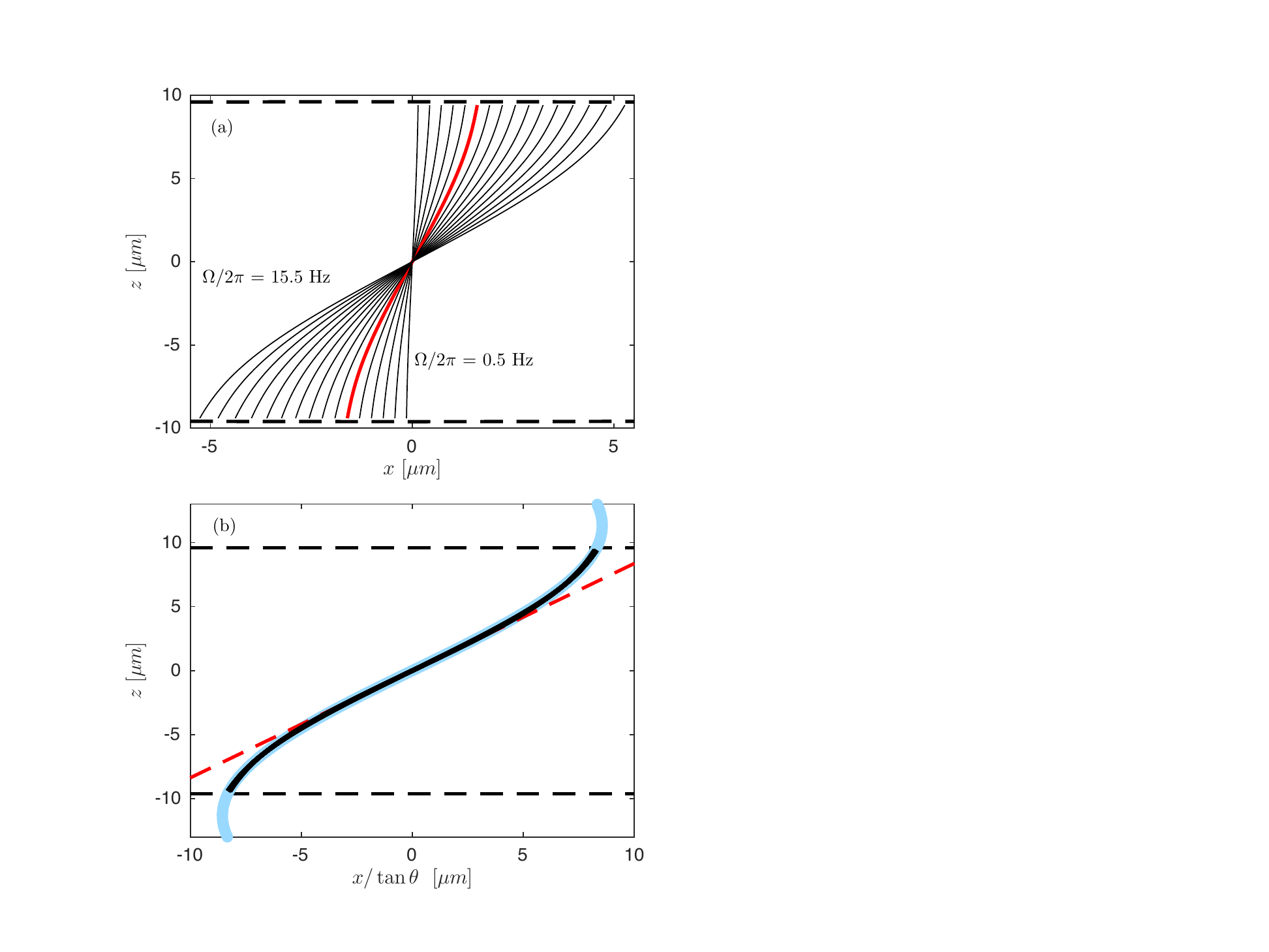}
\caption{(a) Vortex-core profiles of rotating-trap stationary states for a selection of rotation frequencies, each separated by 1 Hz, from $\Omega$ = 0.5 to 15.5 Hz. The thick red line marks the initial state used for the real-time precession dynamics shown in Fig.~\ref{Fig:ExpRes} (a). The dashed lines mark where the Thomas-Fermi approximation predicts the density to vanish. (b) The same vortex-core profiles collapse onto a single curve (black solid lines) when the $x$-axis coordinates of each vortex are rescaled by its central slope, $dz/dx =1/\tan\theta$. The thick blue curve is a sine-function fit (see text) whereas the dashed red line is a straight line, fitted to central slope, to guide the eye. Parameters are the same as in Fig.~\ref{Fig:LnE}.
} 
\label{Fig:TiltProf}
\end{figure}

Next, we investigate how the shape of a tilted solitonic vortex depends on a range of precession frequencies, from $\Omega/2\pi$ =  0.5 Hz to 15.5 Hz, in Fig.~\ref{Fig:TiltProf} (a).
The precession phases are chosen such that the cores lie exclusively in the $y=0$ plane; incidentally, these were also used as initial states for the dynamical simulations displayed in Fig.~\ref{Fig:Precess}.
Remarkably, when the $x$-axis is multiplied by the central slope of the corresponding vortex core, i.e., $x\to x/\tan\theta$, all profiles neatly collapse onto a single curve as shown in Fig.~\ref{Fig:TiltProf} (b).
The tendency for a vortex line to exit its condensate at an angle perpendicular to the surface \cite{Fetter2001} suggests that the most natural deformation, for a tilted ($\theta>0$) solitonic vortex, is a sine function
\begin{equation}
x = D \sin (kz)  \, , \label{sinky}
\end{equation}
with $k=\pi/(2R_\perp)$ and amplitude $D$.
It is worth noting that this deformation is analogous to the lowest Kelvin mode for a vortex of length 2$R_\perp$, but with the distinction that a Kelvin mode in a uniform superfluid is a helix whereas the states considered here lie in a plane.
If one fits Eq.~(\ref{sinky}) to the profiles in Fig.~\ref{Fig:TiltProf}, one finds a value of $k$ slightly smaller than expected, consistent with the fact that the density is not quite zero in the region $|z|>R_\perp$, where it vanishes smoothly. In detail, for the condensate in Fig.~\ref{Fig:TiltProf}, the chemical potential is $\mu = 9.72 \hbar \omega_\perp$ which corresponds to a transverse Thomas-Fermi radius $R_\perp=4.41 a_\perp$, where $a_\perp=\sqrt{\hbar/m\omega_\perp}$. The analytic Thomas-Fermi prediction is then $\pi/(2R_\perp)=0.36/a_\perp$, while a best fit to the GPE data gives $k=0.30/a_\perp$.

Single bent vortices were studied in \cite{Rosenbusch02,Aftalion2003,Bretin2003,Modugno03,Garcia01a,Garcia01b,Komineas2005} for BECs in rotating traps. On the one hand, the so-called S-shape vortices in cigar-shaped condensates with a fast trap rotation about the long axis \cite{Rosenbusch02,Aftalion2003,Bretin2003} can be seen as the high $\Omega_{\rm tr}$ counterpart of our tilted solitonic vortices. In fact, for large $\Omega_{\rm tr}$,  transverse vortices become so stretched that they develop long straight portions, aligned parallel to the trap's long axis, hence losing their solitonic-vortex character. Eventually, as the trap rotation speed increases further, such a vortex continuously evolves to become a perfectly straight line, coaxial with the trap's long axis. Furthermore, the cigar-shaped trap of \cite{Aftalion2003} was not symmetric about the trap's long axis. This means that a vortex stationary state in the rotating frame would not simply precess if evolved in real time with the trap's rotation switched off. On the other hand, in the extremely dilute limit where kinetic energy dominates over interaction energy, Ref.~\cite{Komineas2005} found that instead of a sine-shaped core the tilted ( $\theta >  0$) solitonic vortex remains as an almost rectilinear line. In contrast, our BECs are not dilute, i.e., they are well within the Thomas-Fermi limit, and we consider an axially symmetric nonrotating trap with slowly precessing solitonic-vortices.

\subsection{Extraction Procedure}\label{Sec:ExtProc}

In our experiments we implement forced evaporation to produce sodium BECs of around $2\times 10^7$ atoms in the same cigar-shaped traps as used for our numerics, i.e., with $\omega_\perp = 10\omega_x$ and $\omega_\perp/2\pi = 92$\,Hz. A detailed description of our experimental procedure can be found in Refs.~\cite{Serafini2017,Lamporesi13} while here we highlight the relevant points.
The speed of the temperature quench is controlled such that a given condensate typically inherits one, or a few, solitonic vortices via the Kibble-Zurek mechanism \cite{LamporesiKZM13,Donadello16,Weiler08,Freilich10,Corman14,Navon15,Chomaz15}.
For the purpose of investigating vortex dynamics in real time we utilize an imaging scheme, also presented in Ref.~\cite{Serafini2017}, that periodically probes the condensate in a minimally invasive manner \cite{Freilich10,Ramanathan14}.
A small fraction $\approx$ 1\% is extracted every 12 ms and after expansion this is imaged, leaving the trapped condensate otherwise intact.
The extraction is performed by transferring atoms from the trapped $|F = 1, m_F=-1\rangle$ state to the untrapped $|1, 0\rangle$ state using a radio frequency field. Since only one component feels the magnetic trap, the energy difference (and hence the resonance condition) between the two states is position dependent and enables us to selectively address different spatial regions of the condensate.
The gravitational sag, of 30 $\mu$m in the $z$ direction, is larger than the condensate radius and this allows us to linearly sweep the radio frequency field to produce a single resonance front that travels from top to bottom.
This sweep of the extraction front has the effect of compressing the extracted portion in the vertical direction and enhancing self-interference effects, which aids with the gathering of {\it in situ} information about the position and orientation of the solitonic vortices.
As the extracted fraction expands it also falls under gravity while interacting significantly with the trapped condensate for about 3 ms, after which they become spatially separated. Finally, following a 13ms time of flight (TOF), the extracted portion is imaged.

To better understand how the expansion images from the above procedure relate to the {\it in situ} positions and orientations of the vortices, we perform full numerical simulations using the time-dependent GPE [Eq.~(\ref{Eq:GPE})].
The interactions between trapped atoms, and between untrapped and trapped atoms, are of the same strength and are characterized by the scattering length 54.54(20) $a_0$; the subdominant interactions between extracted atoms have a scattering length of 52.66(40) $a_0$ \cite{Knoop2011}.
On the one hand, the expansion dynamics is relatively fast and this allows us to treat the in-trap vortex positions as fixed during the extraction sweep. On the other hand, the global phase of the trapped condensate continues to evolve and it turns out to be crucial to account for this during the extraction.

As was the case in Sec.~\ref{Sec:Numerics}, the in-trap part of this simulation treats $N=8\times10^5$ sodium atoms in a harmonic trap with the same confinement parameters as for the experiment.
The discrepancy of atom number between theory and experiment corresponds to a chemical potential difference of a factor of three and, hence, Thomas-Fermi radii that differ by a factor of $\sqrt{3}$. It is not feasible for us to simulate the full experimental atom number since, as a reference, producing the results in this paper already consumed around four weeks of computer time on 100 cores. However, due to the findings in Ref.~\cite{Serafini2017} and the comparisons between theory and experiment in this paper, we expect this discrepancy not to be of qualitative importance.
To account for the different Thomas-Fermi radii between theory and experiment, when optimizing the vertical compression of the extracted portion, the numerical extraction sweep is 12 kHz/ms, while it is 10 kHz/ms for the experiment. The numerical results that follow assume a 10 ms TOF before imaging the extracted fraction (cf.~13 ms for the experiment); we have numerically checked that this results in only minor differences.
While the creation of our initial states is described in Sec.~\ref{Sec:Numerics}, during the course of the expansion we interpolate and enlarge the grid such that the one used for the final part of the expansion has size $\{L_x,L_y,L_z\}$ = \{229,139,109\}\,$\mu$m and $\{N_x,N_y,N_z\}$ = \{600,240,500\} points, respectively. We have checked that all results presented here are numerically converged to $\sim$1\% or better.

\subsection{Extraction Results: Experiments and Numerics}\label{Sec:ExtRes}

\begin{figure*}[t!]
\includegraphics[width=0.6\textwidth]{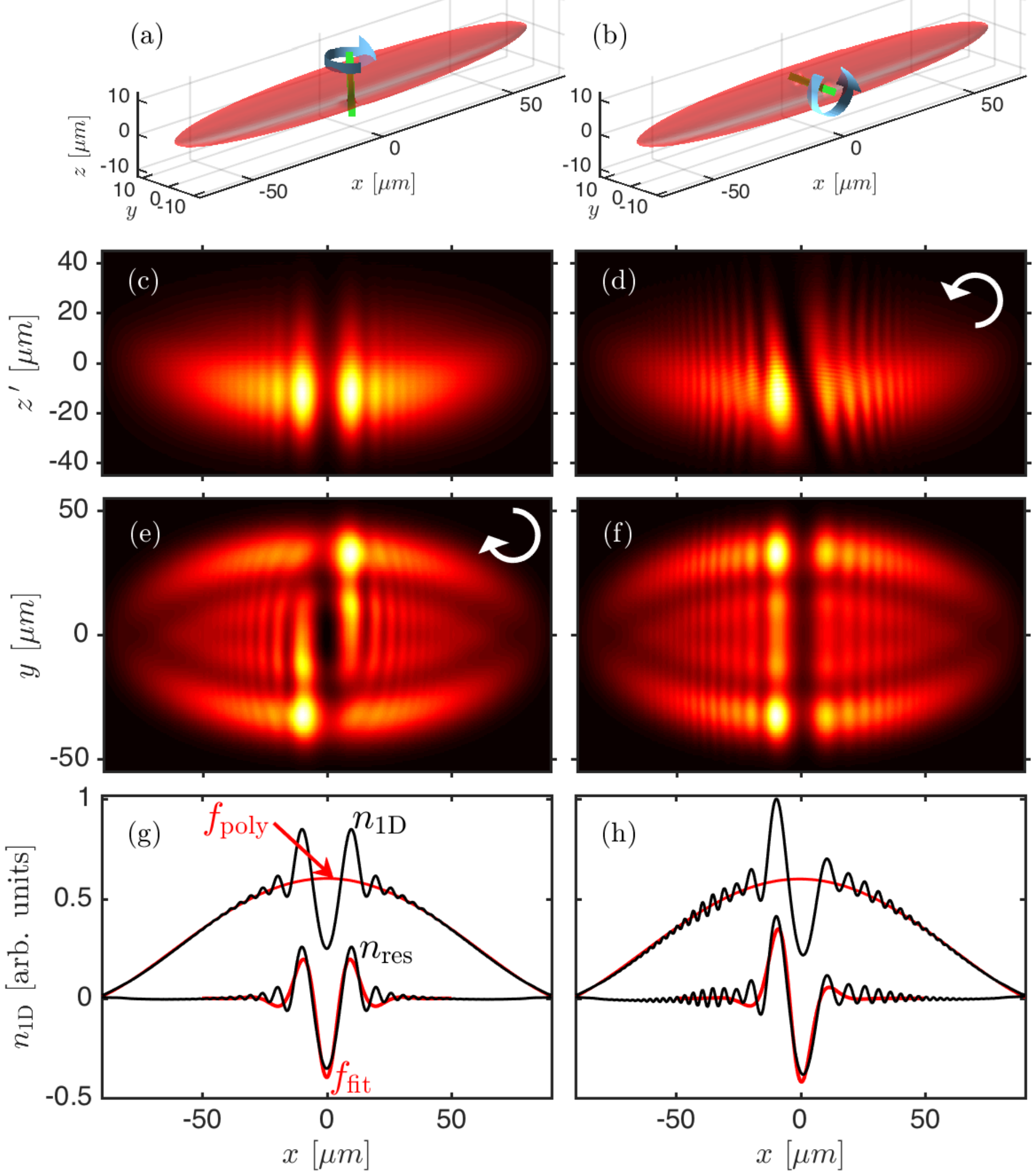}
\caption{Simulation of expansion and image processing for a vertical (left) and a horizontal vortex (right). The directions of superfluid flow are indicated by the semicircle arrows. (a)(b): {\it In situ} density isosurfaces at half of the peak density. (c)-(f): 2D column densities of the extracted portions in the $x-z^\prime$ and $x-y$  planes after a 10ms TOF. Note that $z^\prime=z+z_0$, for constant $z_0$, is used to account for the displacement due to gravity, which acts in the $-z$ direction. (g)(h): The integrated 1D densities $n_{\rm 1D}(x) = \int\int |\psi(\mathbf{x})|^2dydz$ are plotted, along with their polynomial fits $f_{\rm poly}$, the residual $n_{\rm res} = n_{\rm 1D}-f_{\rm poly}$ and the fit to the residual $f_{\rm fit}$ [Eq.~(\ref{Eq:fit})].
For the vertical vortex on the left, fitting Eq.~(\ref{Eq:fit}) gives $\{\delta,x_\nu\} = \{0,0\}$, while for the horizontal vortex on the right $\{\delta,x_\nu\} = \{-1.35,-3.5\mu {\rm m}\}$.
For the in-trap part of the simulation, the parameters are the same as in Fig.~\ref{Fig:LnE}; details of the expansion are provided in the text.
}
\label{Fig:Expns}
\end{figure*}

Numerical simulations relating the {\it in situ} vortex orientations to the corresponding expanded extractions are shown in Fig.~\ref{Fig:Expns} for a vertical (left) and a horizontal (right) vortex.
The compression in the vertical ($z$) direction, evident in Figs.~\ref{Fig:Expns} (c)(d), is remarkable given that this direction would normally see the greatest expansion if the extraction had instead been uniform (not swept).
By considering a top-down view, i.e., the $x$-$y$ plane in Figs.~\ref{Fig:Expns} (e)(f), the extracted fraction is seen to be framed by a high density elliptical border that in turn surrounds an ellipse of low density. This effect is due to the interactions with the trapped condensate, and exists even in the absence of any vortex.

Importantly, though, by contrasting the 2D extracted fractions for the two different vortex orientations, additional regions of constructive and destructive interference are apparent.
The physical processes involved in the creation of these vortex-induced asymmetries are complicated but two important contributions are as follows. (i) The vortical superfluid velocity is greatest when forced to flow near a boundary and, after expansion, such a region of enhanced velocity tends to leave behind a hole and an adjacent bump \cite{Donadello14,Tylutki15}. This mechanism is important for, e.g., a vertical vortex as can be seen in Fig.~\ref{Fig:Expns} (e) where two high-density bumps are positioned diagonally about the core. (ii) During an extraction sweep, the atoms that are released early experience a drop in potential since they no longer feel the trap. The result is that the wavefunction phase evolves faster for those atoms that remain trapped, and the eventual interference between the atoms released early and those released late causes destructive or constructive interference depending on the {\it in situ} phase pattern about the core. This process was important for the expanded horizontal vortex, seen in Figs.~\ref{Fig:Expns} (d)(f), where constructive interference can be seen on the $-x$ side, but not the $+x$ side.

Since each experiment periodically images a given condensate, it is useful to process the extraction images to determine parameters that keep track of the {\it in situ} vortices.
The first step is to integrate either column-density over the remaining radial direction to obtain the 1D density, $n_{\rm 1D}(x) = \int\int |\psi(\mathbf{x})|^2dydz$, as plotted in Figs.~\ref{Fig:Expns} (g)(h).
A 1D residual density, $n_{\rm res}(x) = n_{\rm 1D}(x)-f_{\rm poly}(x)$, is then obtained by subtracting a fourth-order polynomial fit $f_{\rm poly}(x)$.
The residual is subsequently fitted by the function,
\begin{equation}
f_{\rm fit}(x) = \frac{A\cos[B(x-x_\nu) + \delta]}{\cosh^2[(x-x_\nu)/C]}~, \label{Eq:fit}
\end{equation}
for fit parameters $A<0$, $B$, $\delta$, $C$, $x_\nu$.
It follows, then, that $x_\nu$ provides a measure of the vortex position along the long axis while the phase $\delta$ furnishes a means of tracking its orientation.
For the cases considered in Fig.~\ref{Fig:Expns} we find that for the vertical vortex, $\{\delta,x_\nu\} = \{0,0\}$, while for the horizontal vortex, $\{\delta,x_\nu\} = \{-1.35,-3.5\mu {\rm m}\}$.
Note that for the latter case, the value of $x_\nu$ slightly misrepresents the position of the vortex, which lies in the $x=0$ plane.
Since a horizontal vortex that is oriented in the -$y$ direction has a phase $\delta = -1.35$ then, by symmetry, a horizontal vortex of opposite sense must have $\delta = +1.35$.
An important point is that although this fitting procedure can ascertain that a vortex is vertical, it cannot determine its sense.
However, Fig.~\ref{Fig:Expns} (e) demonstrates that the sense of a vertical vortex can easily be attained from top-down images in the $x$-$y$ plane if one notes the locations of the diagonal high-density bumps about the core.
Furthermore, our simulations (not shown here) illustrate that adding a real-time imaging capability along the vertical direction, which was not feasible in the present experiments, would clearly reveal the $y$ position of off-center vertical vortex cores (see Fig.~\ref{Fig:Expns} (e) for comparison).

\begin{figure}
\includegraphics[width=1\columnwidth]{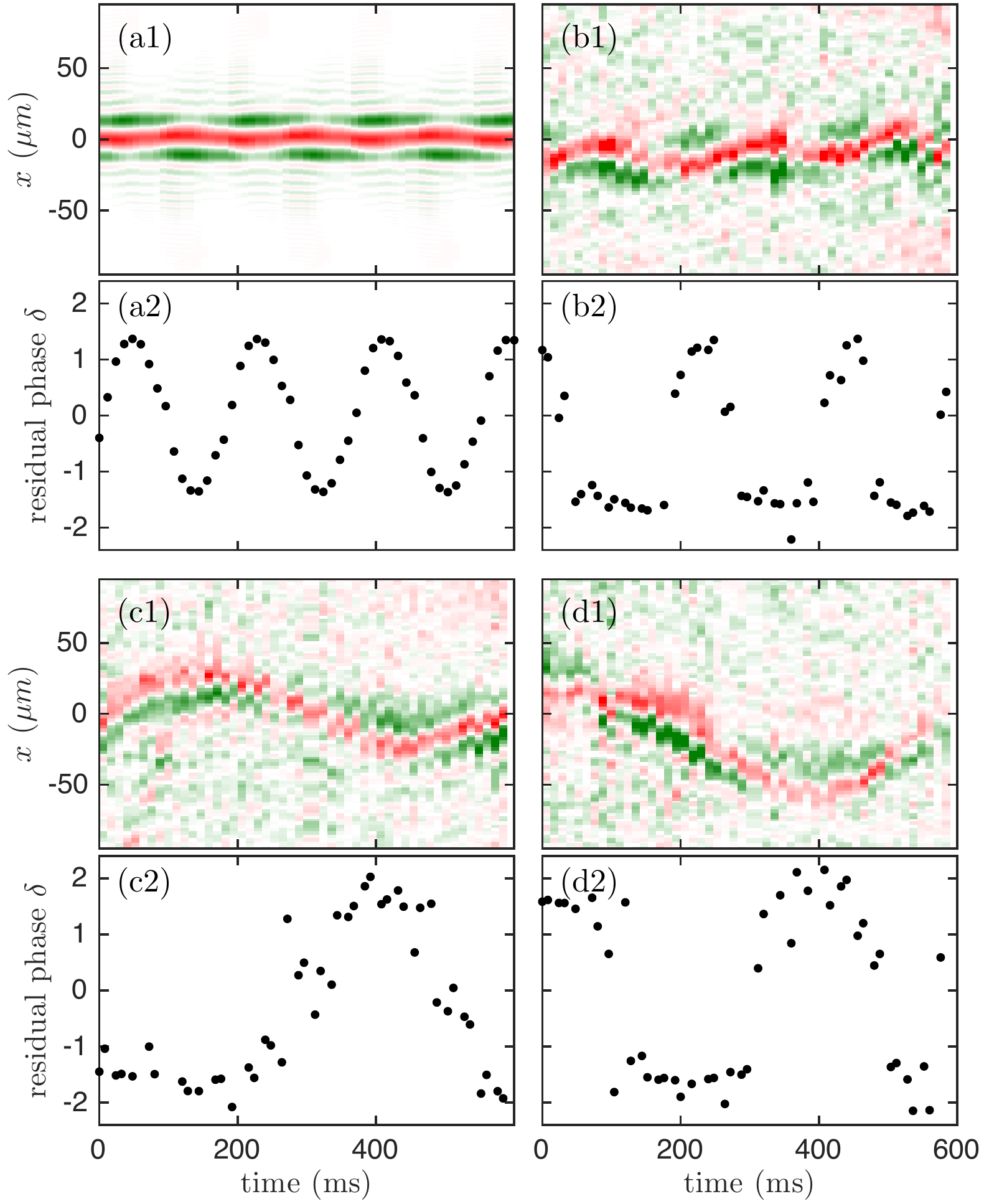}
\caption{(a1)-(d1) Doubly integrated 1D residual densities $n_{\rm res}(x)$ after a partial extraction and expansion and (a2-d2) the corresponding phase of their fits $\delta$ (2), as a function of time. The central (red) color is negative while the outer (green) color is positive. (a) Numerical extractions from an in-trap simulation that used the initial state marked by the thick red line in Fig.~\ref{Fig:TiltProf} (a), having a precession frequency of $\Omega/2\pi=5.5$\,Hz. (b)-(d) A selection of experimental runs. The parameters are described in Sec.~\ref{Sec:ExtProc}.
} 
\label{Fig:ExpRes}
\end{figure}

A numerical calculation of the 1D residual $n_{\rm res}(x)$, as a function of time, is presented for a precessing vortex in Fig.~\ref{Fig:ExpRes} (a1).
The central (red) color represents a negative value while the outer (green) color is positive.
The initial state for this simulation, highlighted as the thick red line in Fig.~\ref{Fig:TiltProf} (a), is evolved according to the time-dependent GPE [Eq.~\ref{Eq:GPE}].
As the vortex precesses, the wavefunction is periodically saved and each of these is then used to initiate an extraction simulation to produce a single time slice of Fig.~\ref{Fig:ExpRes} (a).
This particular vortex has a tilt of $\theta$ = 0.22 radians and a precession frequency $\Omega/2\pi$ = 5.5\,Hz, as can be seen from Fig.~\ref{Fig:Precess}. 
The fitted phase $\delta$ [see Eq.~(\ref{Eq:fit})], shown in Fig.~\ref{Fig:ExpRes} (a2), is a smoothly varying function of time that maps to the {\it in situ} orientation of the precessing vortex.

We present experimental evidence for a precessing solitonic vortex in Fig.~\ref{Fig:ExpRes} (b). This has a precession frequency of $\approx$\,5 Hz, which is very similar to that of our numerical simulation in Fig.~\ref{Fig:ExpRes} (a).
Since the experiment has a chemical potential $\approx$\,3 times larger than in our simulations, the scaling of our analytic model, Eq.~(\ref{Eq:PrecessAna}), suggests that the tilt angle be $\approx 2$ times larger for this experiment, i.e. $\theta \sim 0.4$ radians.
As was the case for the simulation, in Fig.~\ref{Fig:ExpRes} (b2) the residual's phase displays an oscillatory behavior. However, a difference here is that $\delta$ now has an asymmetry, with some saturation near $- \pi/2$.
A possible explanation for this bias is a slight tilt ($\sim 1$ degree) of the imaging camera, which looks down the $y$ axis, effectively rotating the $x$-$z$ plane relative to the direction of gravity. 
To help visualize this, consider the simulation in Fig.~\ref{Fig:Expns} (d), where the density minimum of the vortex core exhibits a tilted, relatively narrow {\it canyon} in the vertical direction.
The sensitivity to a camera tilt is expected to be even more pronounced in the regime of the experiment, for which the healing length is smaller, the Thomas-Fermi radii are larger and the TOF is longer.
Two further experimental examples of precessing vortices are presented in Figs.~\ref{Fig:ExpRes} (c) and (d).
In addition to the precession evidenced by the changing order of colors (and the corresponding oscillations of $\delta$), these vortices orbit about the BEC's origin, manifested here as oscillations of their $x$ coordinate, as they follow contours of constant Thomas-Fermi density \cite{Anderson00}.
We note that it is not possible to directly obtain a vortex's axial tilt from our TOF images, due to complications from the interactions between the extracted portion and the trapped condensate, as well as interference effects within the extracted portion, thus prohibiting a quantitative comparison with our theoretical prediction (\ref{Eq:PrecessAna}) for the precession frequency as a function of the tilt angle.

\begin{figure}
\includegraphics[width=1\columnwidth]{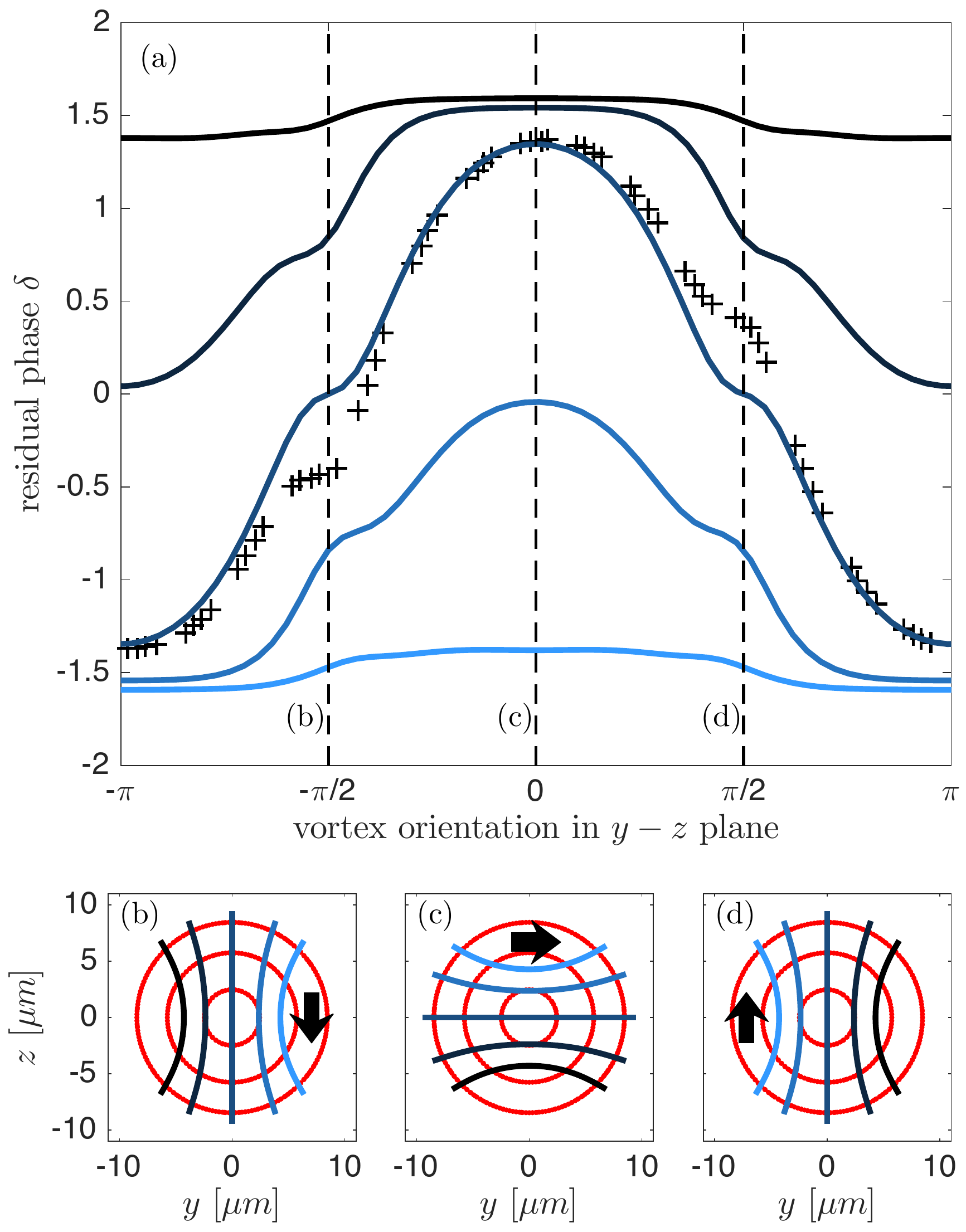}
\caption{(a) Numerically determined residual phase $\delta$ after a 10\,ms partial expansion as a function of \emph{in situ} vortex orientation. Solid curves are for untilted (strictly in the $x=0$ plane) vortices while the `+' symbols are for the tilted precessing vortex already shown in Fig.~\ref{Fig:ExpRes} (a). The middle curve is for a vortex that runs through the origin while the others are for off-center vortices. The three dashed lines mark when the untilted vortices are oriented along the vertical (-$z$), horizontal (+$y$) and vertical (+$z$) direction, and these correspond to the examples in the lower panels that illustrate the positions of the off-center vortices. (b)-(d) Cuts through the $x=0$ plane illustrating the positions of the vortex cores in the main plot. Density isocontours are at 10\%, 50\% and 90\% of the peak density. The arrows indicate the sense of vorticity. Note that the curves in the upper and lower subplots have a matching color code to guide the eye.
} 
\label{Fig:phase_orient}
\end{figure}

An intriguing question is how the relationship between the residual and vortex orientation is modified for off-center vortices, i.e., those that do not pass through the $x$ axis.
The residual phase versus the {\it in situ} orientation angle is plotted for various off-center vortices in Fig.~\ref{Fig:phase_orient} (a).
These untilted vortices, which lie in the $x$ = 0 plane, are represented as solid lines that vary in color from black to light blue (gray).
To aid with their visualization, the positions of these vortex cores are plotted for three angles (the three vertical dashed lines in the main plot) in Figs.~\ref{Fig:phase_orient} (b)(c)(d), with a matching color code.
In the main plot, the stationary-state vortex (the straight line in the lower panels) has a $\delta$ that is symmetric about zero, as expected.
For comparisson, we collapse the precessing vortex data from Fig.~\ref{Fig:ExpRes} (a2) onto the main plot of Fig.~\ref{Fig:phase_orient}, and mark these with plus symbols. The behavior is fairly similar to that of the stationary-state vortex, indicating that the axial tilt ($\theta>0$) itself does not have a significant effect on the residual.
The behavior changes radically, however, if a vortex is off-center. In particular, the values of $\delta$ become asymmetric and tend to become, either more negative or positive depending on the sense of the vortex relative to its closest boundary.
For the most-off-center vortices, take the one indicated by the black line for example, changing the orientation angle has little effect on $\delta$. Consequently, as a vortex becomes more off-center, $\delta$ is no longer a useful indictor of a vortex's orientation, but instead conveys the sense of the vortex relative to its closest boundary.
As a final note, we can deduce that this off-center-vortex effect is not responsible for the $\delta<0$ asymmetry in Fig.~\ref{Fig:ExpRes} (b2). This is because such an off-center vortex would also orbit about the condensate center \cite{Anderson00}, which would be evident as large oscillations of the vortex position along the $x$ axis, contrary to observations in Fig.~\ref{Fig:ExpRes} (b1).

\section{Conclusions}\label{Sec:Conc}

Solitonic vortices are highly-localized objects, both in terms of their energy and angular momentum densities, in stark contrast to vortices of untrapped systems.
With this as motivation, we developed a theoretical model that treats solitonic vortices on a similar footing to classical spinning tops.
Using our minimally destructive imaging scheme we experimentally observed this spinning-top behavior by periodically imaging a given condensate in real time.
We performed 3D Gross-Pitaevskii simulations to investigate how the precession frequency varies as a function of the axial tilt angle, and comparisons of these with our analytic prediction further supported our spinning-top model, while also quantifying its limitations.
Finally, we carried out Gross-Pitaevskii simulations of our experimental extraction and imaging scheme. From these we suggested improvements, such as the addition of real-time imaging capability along the vertical ($z$) direction to keep track of the sense and radial position of vortices when they are vertical. Our simulations also demonstrated how to interpret expansion images when vortices are off-center, i.e.~when they do not pass through the $x$ axis.

An interesting question arises regarding the role of a trap asymmetry in the radial plane or, in other words, the squashing of the cigar-shaped trap along one of its short directions. For small radial asymmetries the vortex is still expected to precess about the trap's long axis, albeit with a small periodic change of the tilt angle to preserve its length and hence conserve its energy. On the other hand, vortex dynamics for a highly squashed cigar becomes dominated by the tight direction, a situation for which a tilted vortex exhibits a precession about this new axis \cite{Powis2014}. The nature of the transition between these two orthogonal precession axes, as a function of the radial asymmetry, presents an intriguing consideration for future research.\\

{\bf Acknowledgments}
We thank Luca Galantucci,  Carlo Barenghi, Nick Proukakis and Lev Pitaevskii for useful discussions.
We acknowledge the EU QUIC project and Provincia Autonoma di Trento for financial support.\\


%

\end{document}